\documentclass[letterpaper, 10 pt, conference]{ieeeconf}  

\IEEEoverridecommandlockouts                              
\usepackage[colorinlistoftodos]{todonotes}

\usepackage{graphics} 
\usepackage{epsfig} 
\usepackage{color}
\usepackage{soul}
\usepackage{float}
\usepackage[switch]{lineno}

\title{\LARGE \bf
Compact Convolutional Neural Networks for Multi-Class, Personalised, Closed-Loop EEG-BCI}

\author{Pablo Ortega$^{*}$, C{\'e}dric Colas$^{*}$ \& A. Aldo Faisal
\thanks{P.O., C.C. and A.A.F. are with the Brain \& Behaviour Lab, Dept. of Bioengineering \& Dept. of Computing, Imperial College London, UK. C.C.'s present address is French Institute for Research in Computer Science and Automation (INRIA), Bordeaux, France. {\small Address for correspondence: a.faisal@imperial.ac.uk}}%
\thanks{$^*$ Both authors contributed equally to this work.}}

\begin{document}

\maketitle
\thispagestyle{empty}
\pagestyle{empty}

\begin{abstract}
For many people suffering from motor disabilities, assistive devices controlled with only brain activity are the only way to interact with their environment \cite{wolpaw2002brain}.
Natural tasks often require different kinds of interactions, involving different controllers the user should be able to select in a self-paced way.
We developed a Brain-Computer Interface (BCI) allowing users to switch between four control modes in a self-paced way in real-time.
Since the system is devised to be used in domestic environments in a user-friendly way, we selected non-invasive electroencephalographic (EEG) signals and convolutional neural networks (CNNs), known for their ability to find the optimal features in classification tasks.
We tested our system using the Cybathlon BCI computer game, which embodies all the challenges inherent to real-time control.
Our preliminary results show that an efficient architecture (SmallNet), with only one convolutional layer, can classify 4 mental activities chosen by the user.
The BCI system is run and validated online. It is kept up-to-date through the use of newly collected signals along playing, reaching an online accuracy of $47.6\%$ where most approaches only report results obtained offline.
We found that models trained with data collected online better predicted the behaviour of the system in real-time. This suggests that similar (CNN based) offline classifying methods found in the literature might experience a drop in performance when applied online. 
Compared to our previous decoder of physiological signals relying on blinks, we increased by a factor 2 the amount of states among which the user can transit, bringing the opportunity for finer control of specific subtasks composing natural grasping in a self-paced way.
Our results are comparable to those showed at the Cybathlon's BCI Race but further improvements on accuracy are required.
\end{abstract}

\begin{figure}[b]
	\centering
    \includegraphics[scale=0.25]{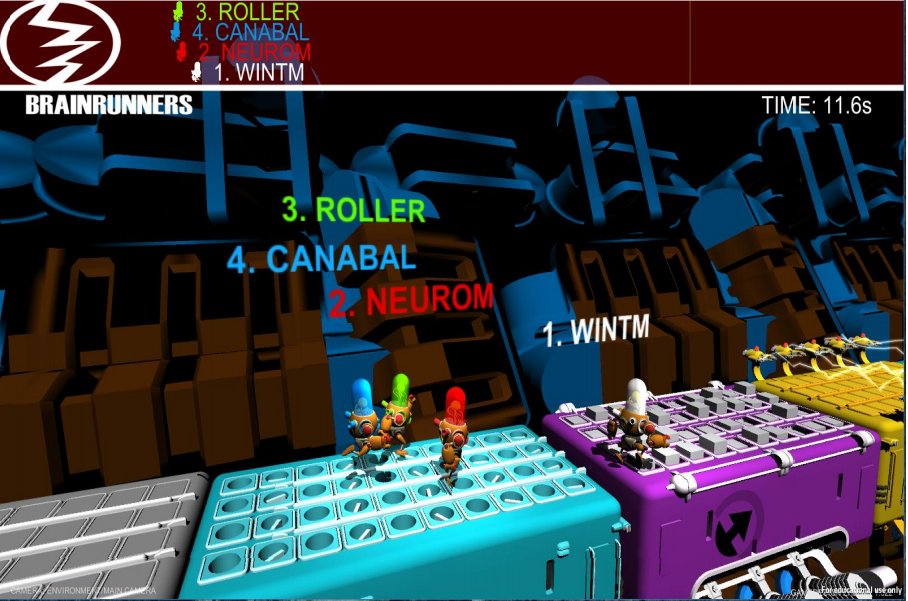}
    \caption{Snapshot from pilot's point of view of the Cybathlon 2016 BrainRunners video-game. Each avatar corresponds to a user competing in the race. Each obstacle is indicated by a different colour. The control is achieved by decoding different mental tasks associated with each desired command. Source: \textit{Cybathlon BCI Race 2016}.}
    \label{fig:brainrunners}
\end{figure}

\section{INTRODUCTION}

BCI devices for paralysed users are typically based on electroencephalographic (EEG) recordings from the scalp. 
While invasive approaches celebrated several successes in the area of closed-loop control, the medical risks, the high clinical intervention, the limited lifetime of implanted electrodes before they have to be surgically removed again and the very high proportion of \emph{a priori} excluded patients place currently severe limits on the practical feasibility of implanted neural interfaces and the use of wearable devices powered by them \cite{makin2017neurocognitive}. 
In the non-invasive domain many EEG decoding approaches have been mainly implemented in offline classification benchmarks \cite{ang2012filter,bashivan2015,tabar2016,lawhern206,schirrmeister2017}. These,however, do not capture the closed-loop nature of control, namely that actions (or failures to act) have consequences that will alter brain activity patterns and events being controlled. On the other hand closed-loop systems in clinical neuroengineering \cite{myrden2015,burget} settings operate with customised systems and dedicated patients but are difficult to reproduce and can only be compared offline. 
Furthermore, BCI setups are usually fine-tuned to very specific tasks, equipment, pre-processing strategies, end-users, etc. that are difficult to compare with each other.
These differences hinder the objective comparison across BCI systems, algorithms and approaches. 
The Cybathlon 2016 BCI race defined a unifying set of rules, conditions and task, the closed-loop BrainRunners game (\cite{riener2014}, see also Figure \ref{fig:brainrunners}),  which addresses several of these comparability shortcomings.
In this game, an avatar runs on a track, faces four different obstacles and the users must respond by a matching action decoded from their brain signals. 
In EEG-BCI this typically involves some form of defined mental activities, imagining movements (motor imagery) or more general thought patterns \cite{bos2011study}. 
The decoding accuracy of the BCI determines the avatar's speed across the track, as wrong or no commands will cause the character to slow down. 
The Brainrunners game exemplifies the features of a useful BCI and also what makes BCI hard: it is multi-class, runs in real-time closed-loop and the rules enforced exclusion of eye or other muscle movement artefacts being misused as BCI signal. The completion time for perfect decoding (i.e. appropriate action chosen at any moment) is about $60$s (depending on the random initialisation of the race track). However, the average completion time  in the public BCI race competition held at the Kloten Arena in October 2016  (Kloten, Switzerland) was  $151$s and with a standard deviation of $27$s, exemplifying the challenge to current EEG-BCI.

Our SmallNet system is devised to be used by people with a broad spectrum of motor disabilities.
We chose Convolutional Neural Networks (CNNs) to perform the feature learning and make it personalised for a given user.  
Moreover, rather than fixing the specific mental activities and fine-tuning the design of a classifier for those particular activities and EEG features, these architectures have been shown to extract good features from diverse offline EEG data. 
In  \cite{tabar2016} common spatial patterns (CSP) features are fed to a CNN combined with stacked autoencoders.
Recent work by \cite{lawhern206} and \cite{schirrmeister2017} showed the efficiency of deeper feature learning by CNNs by directly operating on the raw EEG time-series, instead of EEG standard features. 
Their networks have 4 and 2 convolutional neural network layers. 
This offline implementation \cite{schirrmeister2017} has been successfully applied in online robotic EEG-BCI control \cite{burget}.
Excluding this, most of these approaches are limited to offline analyses over standard BCI data sets.
 
The common perceived disadvantage of CNNs is their lack of data efficiency \cite{ferrante2015data}, i.e. that they have often thousands of free parameters and require very large amount of data, which is why they are typically not regarded as efficient approaches for real-world application in EEG-BCI. 
Therefore a user would have to generate large amounts of user data to train these networks (hours of EEG recordings) making it inefficient and potentially unsuitable for \emph{ad hoc} use such as at a BCI competition or in daily life.

Our aim was to build a simple and user-friendly system that requires a short setup and only about 20 minutes of sample recordings to set up an end-user classifier.  

\vspace{-2mm}
\section{METHODS}
\label{sec:extraction}
\begin{figure}[t]
	\centering
	\includegraphics[width=0.5\textwidth]{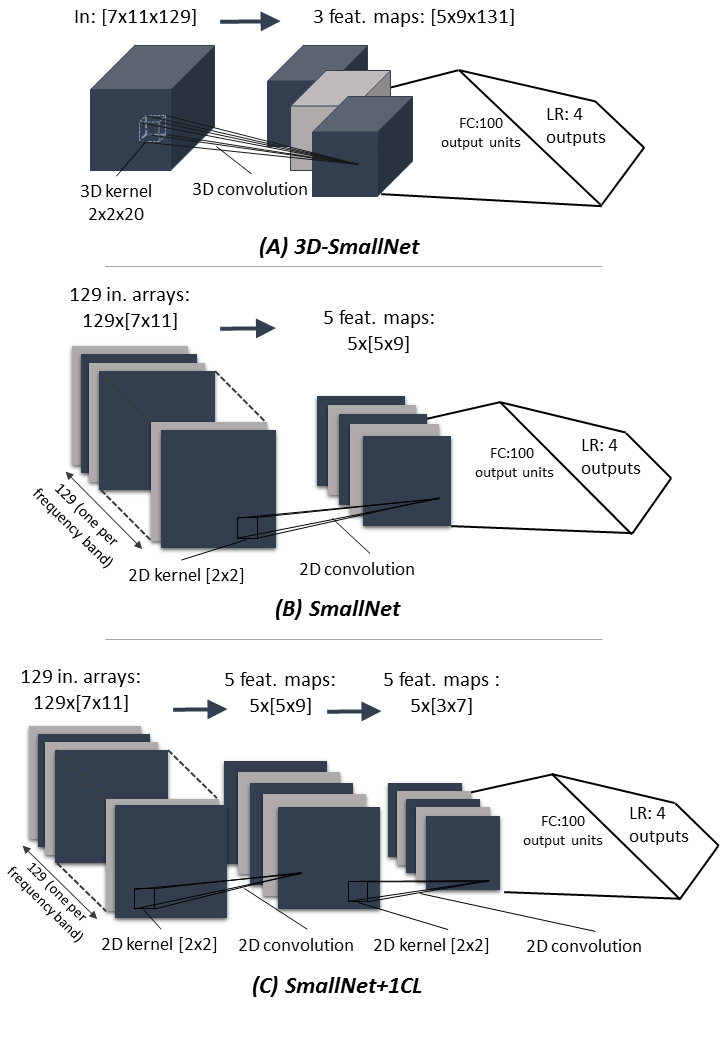}
    \caption{Three different CNNs architectures: 
A 3D-SmallNet (A) using 3D convolution instead of the 2D convolution used by SmallNet (B). 
A convolutional layer was added to Small-Net (C, SmallNet+1CL) and also a fully connected after the first convolutional one,  not presented in the figure (SmallNet+1FC).\vspace{-2mm}}
    \label{fig:archs}
\end{figure}

\paragraph{General approach}
We demonstrate a novel CNN architecture, SmallNet, that requires fewer neuronal layers than other CNN-BCI systems. 
We investigated different CNN architectures that led us to select an efficient one ---SmallNet--- made of one convolutional layer, one fully connected layer and a logistic regression classification layer (see Figure \ref{fig:archs}). 
SmallNet is flexible enough to perform online classification of 4 mental activities that were preferred by a naive user, and complies with Cybathlon's rules, in particular, \emph{a priori} artefact correction.
To overcome the reduced abstraction capabilities of our shallower network architecture, we fed the network with standard EEG features computed from the raw  signal. 
In particular, we used Welch's periodogram features preserving the spatial arrangement of electrodes. Second, we developed a personalisation protocol to select $4$ mental tasks among a set of $8$, based on the performance of the user and his personal preferences.
Third, based on our results from previous steps, we carried out adaptive training of the CNNs and validated the system online in closed-loop.

In the following we describe the three stages of our methodological approach to CNN-based BCI.
We first describe the procedure for data collection and pre-processing, before detailing the choice of architectures, mental tasks, and the implementation of adaptive training. 
Results are then presented before being discussed.

\paragraph{Data collection \& Subject description}
EEG data was recorded using 64 electrodes positioned according to the 10-20 system (reference `Fpz'), using a BrainVision ActiChamp (v. 1.20.0801) recorder.
We used a sampling rate of $500$Hz, a high-pass filter ($0.1$Hz) and a notch filter $50$Hz.
Electrooculogram (EOG) activity was recorded on the right eye to correct for ocular artefacts using independent component analysis (ICA) (Python \textit{MNE} implementation \cite{mne2}).
ICA matrices were computed offline and the rows corresponding to EOG components were removed from the matrix.
Applying this transformation matrix allowed to clean online EEG data from ocular artefacts.

 Note that for technical reasons the data shown is not the Cybathlon pilot. Instead a 28-year old, right handed, naive to BCI  subject volunteered throughout all the stages.
All the experiments were approved by the Imperial College Ethical Committee.

\begin{figure}[b]
	\centering	\includegraphics[width=0.4\textwidth]{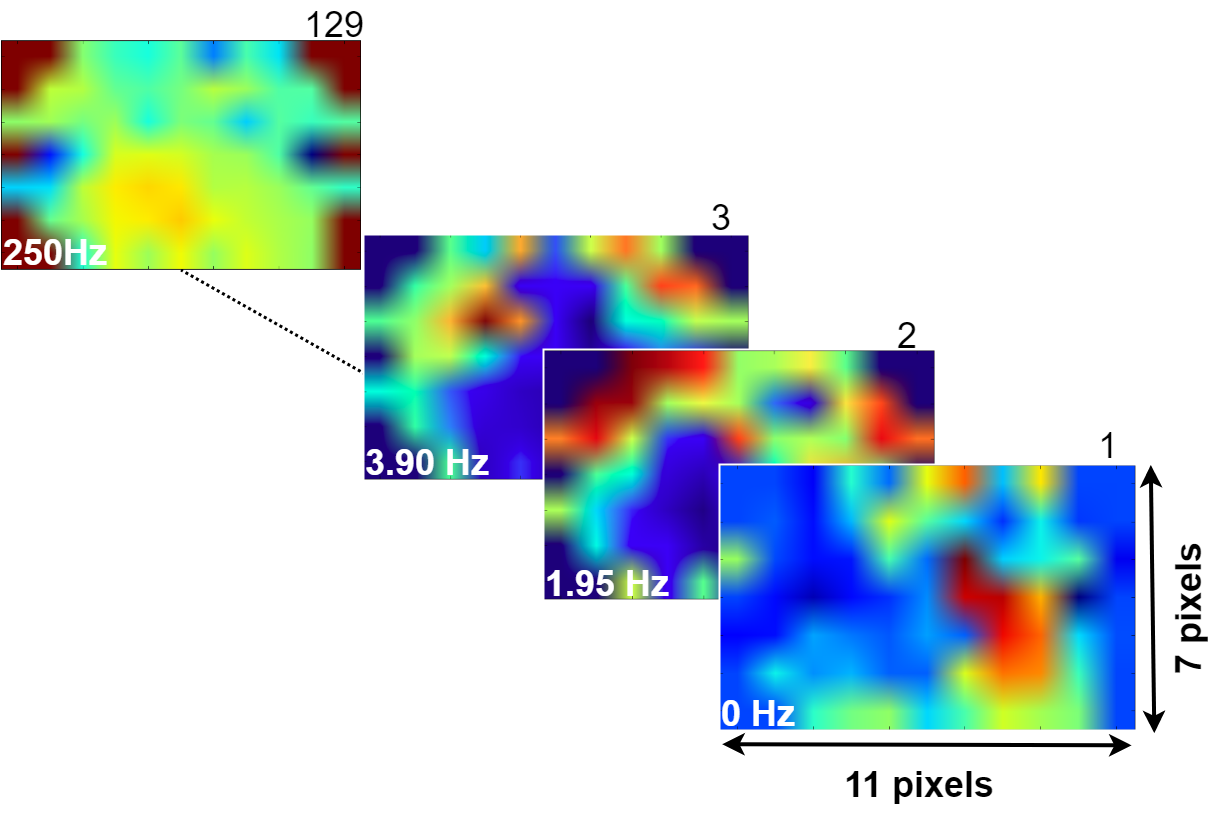}
    \caption{Input features organised in a tensor ($129\times 7 \times 11$) for CNN processing, composed of 129 spectral power images organized as 2D projections of the EEG electrodes topography from the on-line data stream.}
    \label{fig:features}
\end{figure}

\paragraph{Data pre-processing}
Raw EEG signals were pre-processed into a topographical arrangement of the EEG's power spectrum (Figure \ref{fig:features}).
For each EEG channel, a $1.2$s chunk of raw EEG data is split into $75\%$ overlapping segments of length $300$ms.  Each segment is windowed (Hamming window), before its periodogram is computed using the discrete Fourier transform (its squared magnitude).
The power spectrum of the EEG chunk is then estimated by the average of the segments periodograms, this is the Welch's estimate. 
Periodograms span 129 discrete frequencies ranging from $0$ to $250$Hz.
For each frequency, the 64 power values corresponding to the 64 electrodes are projected from the 3D EEG topography to a 2D image.
Considering all $129$ frequencies, we end up with an input tensor of size ($129,7,11$), the empty pixels being filled by extrapolation.
The algorithm is fed by such tensors every $300$ms, which means that any raw EEG chunk used to generate an input tensor ($1.2$s) overlaps with its predecessor at $75\%$.

CNN architectures were implemented in Theano \cite{theano} and the system was built in an Intel i7-6700 CPU at 3.40GHz with an NVIDIA 1080GTX GPU.

\paragraph{Architecture selection - Stage 1}
The tested architectures are presented in Figure \ref{fig:archs}.
Different complexities (building up from SmallNet, Figure \ref{fig:archs} B) of CNNs were tested with the intention of finding one able to abstract enough information from our limited set of examples.
For each run, the weights were randomly initialised following a uniform distribution within the $[-1,1]$ range.

All the architectures were trained using a dataset acquired as follows.
The user watched 20 videos of the game for which the pads and their transitions were uniformly distributed.
At each pad, he was instructed to perform the corresponding motor imageries for the whole duration of the pad (contraction of the feet, stomach, right hand or left hand). 
This setup ensured that training examples were extracted in conditions as close as possible to the online setting.
The raw EEG data is then converted into around $9000$ training examples as explained in Section \ref{sec:extraction}.d.
We performed 5-fold cross-validation, splitting data in 5 chronological segments and randomising them so as to avoid overlaps between training and testing examples (and therefore overfitting).

\paragraph{Mental tasks evaluation - Stage 2}
Using our most efficient architecture, we looked for the combination of four mental tasks that would provide the best classification results.
$8$ mental tasks were preselected: motor imageries (right hand \textit{RH}, left hand \textit{LH}, lips, stomach or feet contraction), higher cognitive processes (mental humming and arithmetic \textit{numbers}) and relaxing.
An experimental paradigm, with screened instructions instead of videos, was used to acquire $100$ times $1$-second EEG segments for each mental activity. These segments were preprocessed and used to train $70$ ($8$-choose-$4$) models, compared with $5$-fold cross-validation.

\paragraph{Adaptive training design - Stage 3}
The last stage consists in the validation and analysis of the performance in real-time conditions and investigates an adaptive training methodology that provides feedback to the user.
During adaptive training, the model is tuned after each race, using up to $2000$ examples from previous races.
This limitation prevents long training times.
The importance of this kind of adaptation has been already addressed in \cite{myrden2015}.

\begin{figure}[b]
	\centering
	\includegraphics[width=0.5\textwidth]{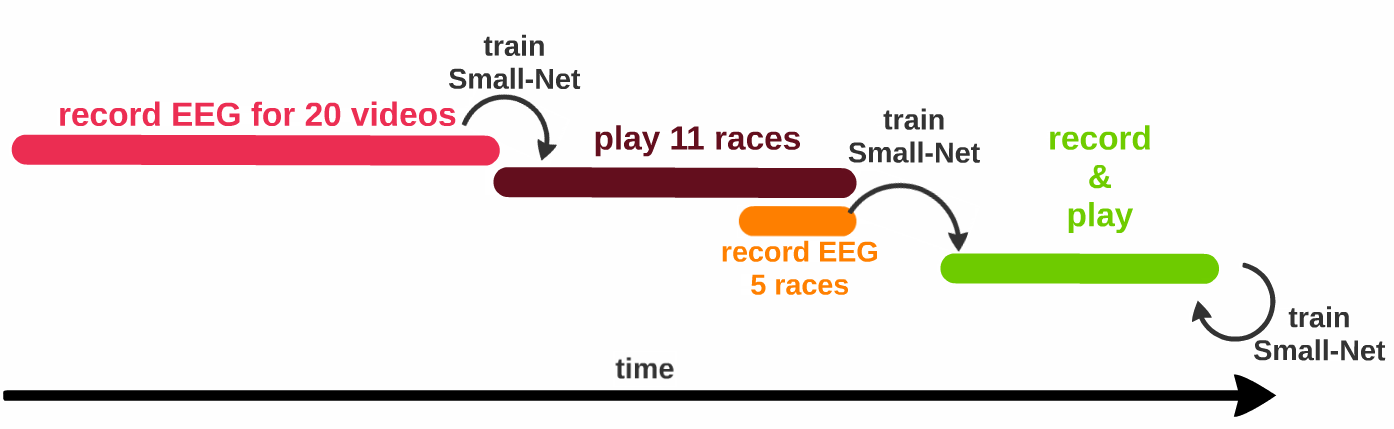}
    \caption[Training strategies]{Training protocol. Non-adaptive (warm colours) and adaptive training (green) strategies. At the beginning of the session, EEG activities are recorded during 20 videos. These are used to train the SmallNet, before this model is used to warm-up the user with actual playing for 11 races. The EEG data recorded during the 5 last races is then used to retrain the model based on online signals. Then starts the adaptive training: after each race, the data recorded is added to the previous dataset and the model is retrained. Note that we limit the number of examples to $2000$ to limit training times in between the races.}
    \label{fig:taining_strat}
\end{figure}

To evaluate online performance, we followed the protocol described in Figure \ref{fig:taining_strat}.
Both the race completion time, and two decoding accuracies were used to evaluate the results.
For the first decoding accuracy ($acc_{1}$), the true label is the one corresponding to the pad where the EEG data was generated (beginning of the time window). For the second ($acc_{2}$), the true label is the one corresponding to the pad on which the avatar stands at the time the action is sent to the game (end of time window).
They could differ if a long decoding time prevented the label predicted from one EEG chunk to arrive on the corresponding pad.

\section{RESULTS}
\paragraph{Architecture selection - Stage 1}
A first analysis discarded 3D-SmallNet due to its long training time, making it impractical for an adaptive online approach.
For the rest of the architectures, the model was trained $5$ times to average out the effect of the network initialisation (random seeds), using a 5-fold cross-validation strategy (Table \ref{tab:arch}).
In the same table results from other works can be seen along with their modality (online or offline) and the number of classes (n. of classes) they classified.
Because there were no clear benefits of adding more layers, we used SmallNet as it required the shortest training time. All statistical tests use a significance level of $p=0.01$.

\begin{table}[b]
\vspace{-2mm}
\centering
\caption{\footnotesize Test accuracy (TA) of relevant architectures.}
\footnotesize
\begin{tabular}{|c|c|c|c|}
\hline
$mt_4$        			 & modality  & n. of classes & $TA_{avg \pm std} (\%)$\\ \hline
\hline
\cite{schirrmeister2017} & offline   & 4			 & $84$    				  \\
\cite{lawhern206}		 & offline   & 4             & $68$                   \\
SmallNet+1CL  			 & offline	 & 4			 & $46.5 \pm 8.6$         \\ 
SmallNet      			 & offline	 & 4			 & $41.8 \pm 1.3$         \\   
SmallNet+1FC  		     & offline	 & 4			 & $41.1 \pm 1.9$         \\ 
\cite{burget} 			 & online    & 5			 & $76.7 \pm 9.1$         \\ 
SmallNet      			 & online	 & 4			 & $47.6 \pm 6.6$         \\ \hline
\end{tabular}
\label{tab:arch}
\end{table}

\paragraph{Mental tasks evaluation - Stage 2}
A Kruskall-Wallis test was used to test whether a given set of mental tasks showed better classification accuracies than another. Table \ref{tab:pairwise_img} shows the number of other sets of mental tasks significantly outperformed by a given set (\textit{n. sig. diff}). Tests are corrected for multiple comparisons using Bonferroni correction. Since there was no absolute best combination of mental imageries (no combination is better than all others), we gave the choice to the user: \emph{RH-feet-relax-mental humming}. This respects the objective of a user-friendly design without significantly hindering the decoding accuracy. 
The chosen combination still presents significant advantages over 3 groups and is not statistically different from any combination above it ($TA_{mean}=46.85\%$ for the selected group compared to $55.01\%$ for group 1 in Table \ref{tab:pairwise_img}).

\begin{figure}[b]
	\centering
    \includegraphics[width=0.4\textwidth]{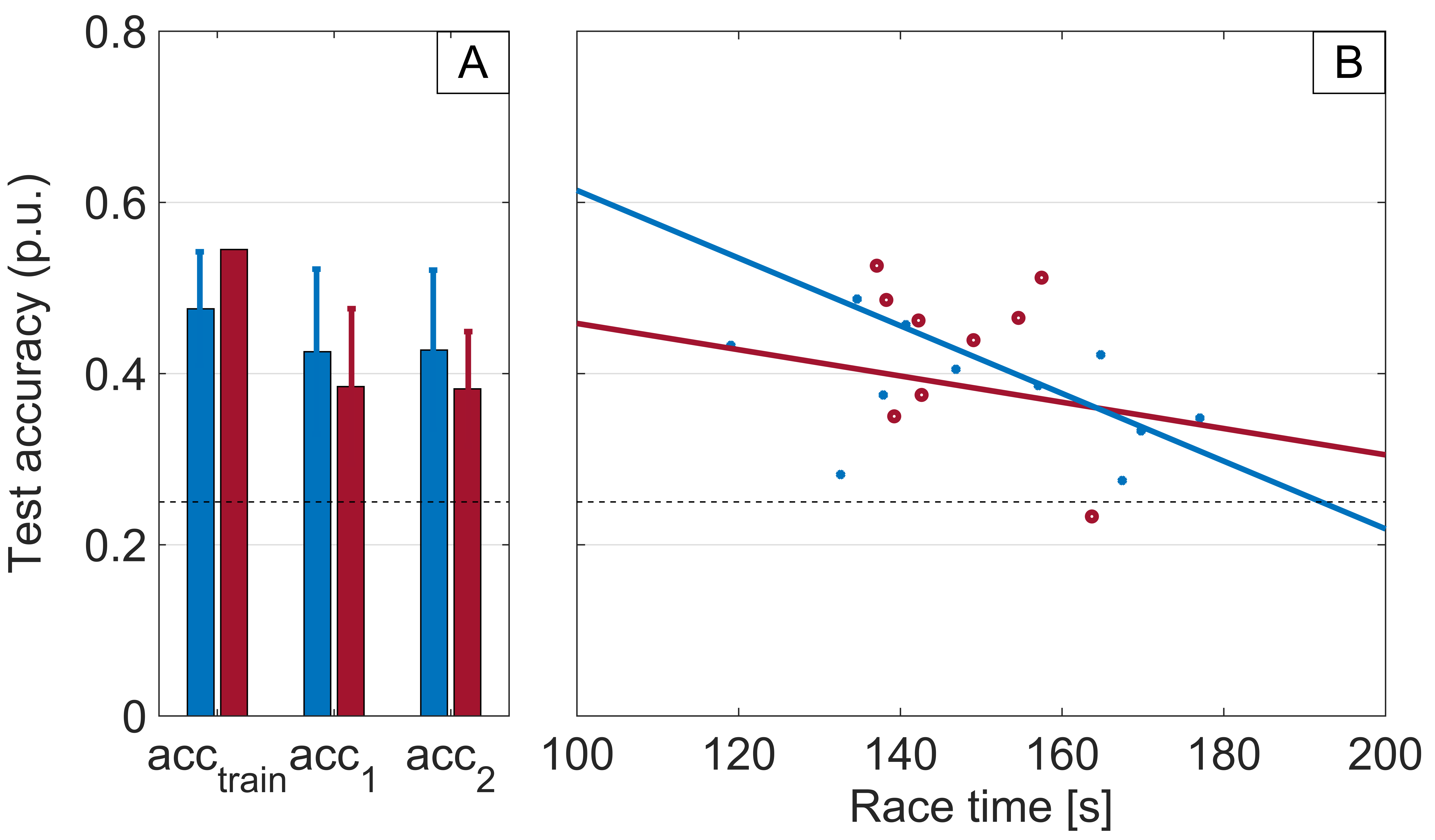}
    \caption{\textbf{A.} Decoding accuracies during play for the model trained with offline video-recorded data (non-adaptive, in red) and trained with recent game-recorded data (adaptive, in blue). $Acc_{train}$ is the accuracy computed offline after training, on the test sets.
    \textbf{B.} Test accuracy as a function of race completion time. Data points and linear fit for the model trained with offline (red) and online (blue) data.Decoding accuracy ($acc_2$) during the game and the time required to finish the race. In both figures, a horizontal black dashed line represents the $0.25$ threshold of random choice.\vspace{-2mm}}
    \label{fig:acc_val}
\end{figure}

\paragraph{Adaptive training - Stage 3}
Figure \ref{fig:acc_val} A shows the decoding accuracy for the online session. There is no statistical difference between the two methods to measure accuracy $acc_1$ and $acc_2$. This means that the decoding time is small enough to decode the EEG chunk into a command before the pad changes. 

The $acc_2$ accuracy computed on the test sets using $5$-fold cross-validation correlates to the race completion time, whether it uses game-recorded ($r=-0.40$) or video-recorded ($r=-0.42$) training data  (Figure \ref{fig:acc_val} B).
In both cases we cannot reject the null hypothesis that the correlation coefficient is lower than zero ($p=0.10$ and $p=0.14$, respectively) indicating that greater accuracies led to shorter race times similarly in both cases.
However, the test accuracy of the model trained on game-recorded data predicted the online decoding accuracy much better than the test accuracy of the model trained on video-recorded data. 
While test accuracies obtained by cross-validation after training the model with recent game-recorded data ($acc_{test}=0.476$) were not significantly different from the decoding accuracy achieved online, the test accuracies of the model trained from video-recorded data (offline training) were systematically higher than those in game conditions ($p=7.14 \cdot 10^{-5}$).
This supports the claim that classification accuracies reported in offline settings often overestimate the achievable real-time performance. This is precisely why such systems should be tested online, in real use conditions.

\begin{table}[b]
\vspace{-2mm}
\centering
\caption[Pairwise comparisons of imageries]{\footnotesize Pairwise comparisons of imageries combinations (top 19)}
\footnotesize
\begin{tabular}{|c|c|c|c|c|c|}
\hline
ranking & im1 & im2   & im3     & im4     & n. sig. diff. \\ \hline \hline

1  & RH  & feet  & lips    & numbers & $18$            \\ 
2  & RH  & feet  & relax   & numbers & $16$            \\ 
3  & RH  & feet  & relax   & lips    & $16$            \\ 
4  & RH  & relax & lips    & stomach & $11$            \\ 
5  & RH  & feet  & stomach & numbers & $11$            \\ 
6  & RH  & feet  & m. humming   & numbers & $10$       \\ 
7  & RH  & relax & stomach & numbers & $10$            \\ 
8  & RH  & relax & lips    & numbers & $10$            \\ 
9  & LH  & feet  & relax   & lips    & $8$             \\ 
10 & RH  & LH    & relax   & lips    & $6$             \\ 
18 & LH  & feet  & relax   & stomach & $3$             \\ 
\textbf{19} & \textbf{RH}  & \textbf{feet}  &\textbf{relax}   & \textbf{m. humming}   & \textbf{$3$}             \\ \hline
\end{tabular}
\label{tab:pairwise_img}
\end{table}

\vspace{-2mm}

\section{DISCUSSION}
Our main contribution consists in the design and implementation of a BCI based on a simple but efficient CNN architecture that achieves classification accuracies well above chance levels, on 4 classes, in real-time conditions and flexible enough to allow the user to choose his preferred mental tasks. Our CNN offers a personalisation protocol enabling different users to evaluate and choose from a larger repertoire of mental tasks. The system was used as part of our participation in the Cybathlon 2016. 

\paragraph{Architecture selection - Stage 1}
We have showed the capabilities of an efficient CNN (SmallNet) to distinguish among four different brain activities in a single example basis, achieving accuracies significantly above-chance ($47\%$ in average, $25\%$ chance level).
We tested variations on the number of layers and network geometry departing from a very simple CNN architecture in order to find whether specific increase in complexity could lead to significant higher accuracies. For our particular setup and user, we found that the simplest model called SmallNet provided accuracies similar to more complex ones and selected it for all further evaluation (cf. Table \ref{tab:arch}).

Bashivan et al. \cite{bashivan2015} proposed a deep learning architecture that also used topographical spectral features as input achieving $91.11\%$ accuracy offline. 
However, compared to ours, their architecture would have required longer training times, was not tested online and was applied to a memory paradigm.
Training CNNs depends very much on the initialisation when data is scarce.
In contrast to Schirrmeister et al. \cite{schirrmeister2017}, we did observe certain instabilities in the training and averaged performance measures over several random initialisation seeds in order to get robust estimates.
The same authors found that a shallower architecture of one convolution layer provided similar results than another one with 4 convolution layers, achieving $84\%$ accuracy offline. 
This was followed by a $76.7\%$ accuracy on the online implementation of the combination of the deep and shallow architectures (or hybrid architecture) of the previous authors by Burget et al. \cite{burget}.
In \cite{lawhern206} a $68\%$ accuracy is reported on sensorimotor rhythms using 3 convolution layers offline. 
In the mentioned studies no artefact correction is applied to the input, neural sources are rather checked \emph{a posteriori}.
Cybathlon rules out this possibility requiring an online implementation for artefact correction which could be a reason for our lower performances.
Indeed, our average race completion time ($147$s) is comparable to those published for the Cybathlon's BCI-Race 2016, where races are finished in between $140$ and $180$s as reported in \cite{stathaler2017}.
In particular, offline accuracies without prior ICA correction reached up to $90\%$ average accuracy for our more complex 3D-SmallNet architecture.
We used the same ICA matrix computed on video-recorded data for all studies to make sure that all data was preprocessed equally.
However, the ICA matrix may change along time and should be updated more often.
Recomputing it after each race in the adaptive training means that the model would be training over differently preprocessed examples, which would have had a bigger impact in the learning stability.
In addition the correction of the signal might remove other globally present oscillations happening simultaneously to and/or at the same frequency range of the blinks.

\paragraph{Mental tasks evaluation and features - Stage 2}
In comparison to the above previous studies \cite{bashivan2015,lawhern206,schirrmeister2017,burget}, we let our participants choose the combination of mental tasks that they felt more comfortable with, backing the decision with offline data. 
Our previous experience with SmallNet showed that spectral power features performed better than \emph{raw-time} features.  Broader bandwidths have been previously used as input to neural networks in \cite{bashivan2015}. Spectral energy features in channel space have been classically used to characterise and study brain activities in several frequency bands, as they have been found to enhance statistical differences (see also \cite{ferrante2015data} on data efficiency). In addition, given the high sampling frequency, the computation of such features greatly reduces the input dimension compared to raw signals of the same time length. 
This enabled our architectures to learn more efficiently compared to the use of raw-time features for which we may have needed more convolutional layers to process appropriately.

\paragraph{Adaptive training validation - Stage 3}
We validated the system using the volunteer preferred imageries (\emph{RH-feet-relax-mental humming} in the top-19).
In particular we found that adaptive test accuracies offered a more reliable prediction of validation accuracies during playing.
A reasonable explanation for this is that adaptive training uses EEG brain activity collected in playing conditions, closer to the ones occurring when the model is used to decode, which supports the experience of the online CNN implementation of \cite{schirrmeister2017}.
Conversely, training the model with EEG signals video-recorded so as to pace the mental tasks without receiving feedback, yielded better test accuracies in data sets with more examples but was less representative of the brain activity during actual playing.
We demonstrated that for a CNN-based BCI, adaptive training can achieve similar performance as offline training. 

Our online results for four classes are more comparable to those reported in \cite{friedrich2013}. CSP was used for feature extraction and linear discriminant analysis (LDA) to classify four categories with \emph{a priori} correction of ocular artefacts. 
Here 8 out of 14 participants achieved around $65\%$ of accuracy (the remaining ones achieved lower accuracies) in a more time-locked experiment compared to our user-driven real-time setting.
Another approach using filter-banks and CSP (FBCSP) for feature extraction and LDA  achieved $80\%$ offline and $68\%$ online accuracies but no artefact correction was performed \cite{chou2015low}.
In \cite{schwarz2016brain} using a Cybathlon complying BCI based in CSP-LDA, the performance between offline analysis and first online session dropped from $79.4\%$ to $51.4\%$ ($35.3\%$ drop).
A similar drop exist between \cite{schirrmeister2017} and its online implementation by \cite{burget}, from $84\%$ to $76.7\%$ ($8.7\%$ drop).
However, in this case, the online implementation added one category to the offline implementation and the drop could be explained by the extra class as well.
Altogether, these figures are comparable to our $29.9\%$ drop for our non-adaptive training and $8.8\%$ for the adaptive, showing that the offline VS online drop in accuracy is somewhat generalisable.

\section{CONCLUSION}
We conclude with the 4 key challenges to EEG-BCI and the use of CNNs as decoding technique therein:
\paragraph{Personalisation} Most EEG-BCI systems require or are tested with a fixed set of mental tasks that the user has to perform. Our results show the value of incorporating the personalised selection of mental tasks for each user in the design of the BCI system. There is considerable insight to be gained from evaluating how and why different mental tasks work best for different users. Crucially, we point to work on expanding the set of mental tasks \cite{grossberger2017investigating} and methods for automatically identifying optimal stimuli  for neural activity measurements \cite{lorenz2016automatic}, as we need to find more principled ways of driving EEG-BCI instead of just confronting users with large libraries of potential mental tasks.
\paragraph{Offline vs Online} EEG-BCI evaluation should move from offline evaluations that offer a stability and, ultimately, a form of overfitting that can result in more complex architectures whose impact on real life conditions is rarely discussed, and in the Cybathlon these become essential to able to compete at all.
The wide gap between optimal ($\approx 60s$) and actual performance ($\approx 150s$) for all participants highlights how far behind on-line BCI decoding is compared to performances reported for off-line BCI benchmark performances in the high $90\%$. Unfortunately the latter is often implicitly equated with the former and this mismatch can be misleading to the wider community beyond the BCI field belief in what the capability of EEG-BCIs are and how much more work needs to be done to address these.
\paragraph{Data efficiency} Online systems require for realistic user setup and subsequent use of data-efficient training algorithms \cite{ferrante2015data}. 
While convolutional neural networks are not typically associated with data-efficient training we achieved this by reducing the number of layers in the networks. However, probabilistic approaches that use more principled manners, such as Gaussian Processes Autoregression \cite{xiloyannis2017gaussian}, to deal with the inherent variability of the neural signals, may be more suitable, if their limitations of processing large amounts of data can be overcome.
\paragraph{Stability} Another challenge is the stability of the EEG-BCI between uses, in the case of the Cybathlon, between the races occurring on the day (which involves setting up and taking down the EEG cap between races, physical movement of the wheelchairs, etc.). It is important to understand how trained models can be transferred from one race to another and find whether any variations are caused by changes in the user mental state, any model updates when performing continuous training or differences at the neural interface (cf. \cite{jayaram2016transfer}). In daily life use of EEG-BCI, the automatic processing of contextual information, such as the environment and tasks performed, could be greatly beneficial in neural engineering \cite{gavriel2013wireless,thomik2013real} to reduce the impact of these variations through data fusion \cite{xiloyannis2017gaussian}.

Meeting these challenges will make EEG-BCI more useful in daily life, paving the way towards broader adoption and retention in daily life use \cite{makin2017neurocognitive}. Our findings with SmallNet offer a personalised, compact system that can be set up with potentially many different naive users in less than an hour. 

\addtolength{\textheight}{-12cm}   



\vspace{-1mm}
\section*{ACKNOWLEDGMENTS}
We thank BrainProducts (Berlin, Germany), Team Imperial and Cybathlon for their support. Our special thanks to our Cybathlon pilot T.N. for his collaboration throughout the years and his valuable feedback. We acknowledge the financial support of the EPSRC CDT HiPEDS (Ref. No. EP/L016796/1).  

\vspace{-2.5mm}
\bibliography{bibdata.bbl}
\bibliographystyle{IEEEtran}

\end{document}